\documentclass[twocolumn,narrowdisplay]{elsart5p}

\usepackage{color}
\usepackage{graphicx}
\usepackage[switch]{lineno}

\begin{document}


\begin{frontmatter}

\title{Estimating a cosmic ray detector exposure sky map under the 
hypothesis of seasonal and diurnal effects factorization}

\author{E. M. Santos}$^{1,2}$, 
\ead{emoura@lpnhe.in2p3.fr}
\author{C. Bonifazi}$^{1,2}$ and
\ead{bonifazi@lpnhe.in2p3.fr}
\author{A. Letessier-Selvon}$^1$
\ead{Antoine.Letessier-Selvon@in2p3.fr}

\address{$^1$ Laboratoire de Physique Nucl\'eaire et de 
	   Hautes \'Energies, T33 RdC, 4 place Jussieu, 75252 Paris Cedex 05,
	   France}

\address{$^2$ Centro Brasileiro de Pesquisas F\'{i}sicas, Rua Xavier Sigaud, 150, 
         22290-180, Rio de Janeiro, Brazil}

\begin{abstract}
The measurement of large scale patterns or anisotropies in the arrival direction of high 
energy cosmic rays is an important step towards the understanding of their origin. 
Such measurements rely on an accurate estimation of the detector relative exposure 
in each direction on the sky: the coverage map.  To reach an accuracy on the 
determination of this map below the 1\% level one must properly identify and 
correct for all the environmental effects that may induce variations in the detector 
exposure as a function of time. In an approach, similar to the one used in anti-sidereal 
time analysis, we propose a method to empirically estimate and correct for those effects 
under the hypothesis that seasonal and diurnal variations can be factorized. We tested 
this method using a model ground detector of cosmic ray air showers, whose aperture 
varies due to the dependence of the air shower development on the atmospheric conditions.
\end{abstract}

\begin{keyword}

Large scale anisotropy \sep Coverage map \sep Cosmic rays \sep EAS-extensive air showers

\PACS 95.85.Ry \sep 96.50.sd \sep 98.70.Sa

\end{keyword}

\end{frontmatter}

\section{Introduction}
\label{intro}

Large scale anisotropy studies play an important role in the search for the origin
of the highest energy cosmic rays. While some results \cite{Agasadipole} have favored 
a scenario with a $\sim$4\% dipole pointing towards the Galactic Center and the Cygnus 
region at EeV (1 EeV = $10^{18}$ eV) energies, analyses with Auger data 
\cite{Aglietta:2006ur} 
have shown consistency with isotropy in this region of the sky. A series of large 
scale complementary analyses with the first two years of Auger data was performed in 
\cite{ICRCEric}, focusing on right ascension modulations and carefully accounting for 
systematic effects, showing as well no indication of significant anisotropy at EeV energies. 
At the PeV (1 PeV = $10^{15}$ eV) range, the Rayleigh formalism applied to right ascention 
modulations of the KASCADE experiment also shows no hints of anisotropies, both for light 
and heavy primaries \cite{Antoni:2003jm}.

In the $10^{11}-10^{14}$ eV range, underground muon and Extensive Air Shower 
(EAS) detectors were able to shed some light into the nature of large scale anisotropies at 
these energies \cite{Ambrosio:2002db, Munakata:1997ei, Amenomori:2004bf, nagashima89}. For 
energies $\sim 10^{11}$ eV, explored mainly by shallow underground muon detectors, cosmic 
rays are believed to be modulated by solar magnetic fields, whereas for energies $\sim 10^{14}$ eV, 
dominated by EAS arrays, the primaries will feel essentially the local structure of the galactic 
magnetic field. Amplitudes of right ascention modulations measured in these two energy ranges 
are $\sim 10^{-4}$ and $\sim 10^{-3}$, respectively. An Earth based detector moving through 
an isotropic 
cosmic ray plasma at rest will register an intensity modulation due to the Earth's rotation 
around its axis, the Compton-Getting (CG) effect \cite{CG}. For the movement of the Earth 
around the Sun, the CG effect should give rise to a solar diurnal modulation. In fact, solar 
diurnal variations measured 
at $\sim10$ TeV \cite{Amenomori:2004bf} (1 TeV = $10^{12}$ eV) are consistent with such an 
expectation. In a similar way, in the cosmic ray plasma rest frame, the movement of the whole 
solar system around the galactic center could give rise to a similar CG effect, however, now 
the modulations should be seen in sidereal time. Analysis in the $\sim 300$ TeV range 
\cite{Amenomori:2006bx} showed no such modulations, indicating that at theses energies, cosmic 
rays corotate with the solar system trapped by the local galactic field inside the arm.

In the energy interval $10^{14}$ - $10^{15}$ eV, linking the two ranges discussed in the first and 
second paragraphs of this section, the EAS-TOP collaboration has reported a solar time modulation 
of amplitude $\sim 3\times 10^{-4}$, in very good agreement with the expected value from the CG 
effect \cite{eastop}. Up to $10^{14}$ eV, a clear signal of sidereal modulation at the 
$\sim 3\times 10^{-4}$ level were also seen by this experiment, and above 300 TeV the amplitudes 
are not significant, showing consistency with the Tibet results.

Although one-dimensional methods, such as the first harmonic analysis (in the case of a uniform 
right ascention exposure detector) of the Rayleigh formalism, the differential East-West method 
\cite{nagashima89} and the Fourier transform method of modified times \cite{Billoir:2007nu}, are 
quite powerful in identifying distortions in the right ascention or sidereal time 
distributions, a complete caracterization of large scale anisotropies rely on the accurate 
knowledge of the detector 
relative exposure in all visible directions of the sky, the two-dimensional coverage map. 
The proper estimation of this map is a very challenging experimental task. All 
changes in the local (Earth based) running conditions of the detector that may 
impact on the exposure must be identified and taken into account. A perfect 
coverage map should contain all variations induced by local effects and none of 
those coming from the sky. In such an ideal case, the excess maps, i.e. the ratio 
of the observed signal over the estimated coverage in each direction of the sky, 
will only show the true sky anisotropies.

The exposure in a given direction of the sky will be a function with a geometrical part 
depending on the local detection angles: the zenith angle $\theta$ and the azimuth $\phi$, 
and a part depending on time. Even though our knowledge on extensive air shower development 
through the atmosphere has clearly increased in the last decades, it is still incomplete 
and Monte Carlo approaches, in addition to being computationally demanding, have a range of 
applicability limited to the cases where the accuracy required is not so much lower 
than just a few percent. Therefore, in the calculation of the geometrical and time 
acceptances, methods which can 
use directly the data are very appreciated \cite{Hamilton:2005za,Benjamin}. Obviously, 
such data based methods will 
face us with subtle effects, since the sample from which we are trying to extract the 
exposure, contains, in principle, the sky anisotropy which we wish to later estimate. The 
usual procedure applied to build the exposure map from the data itself is to perform 
some kind of scrambling into the local angles and arrival times 
in order to wash out the possible sky anisotropies and retain only the local features 
\cite{Clay:2003pv}. The 
role played by the time behaviour of the detector is therefore central in calculating the sky 
exposure map since, on the one hand, detector long term variability induces fake anisotropic 
patterns on the sky, while on the other hand, large scale anisotropy can distort the time 
distribution of events registered by a stably running detector. As an example of the former 
effect, we know that a solar modulation superimposed on the top of a seasonal envelope, both 
genuine local weather effects, will give rise to an apparent sidereal modulation. And for the 
latter, it is widely known that any modulation in right ascension will appear as a sidereal 
variation in the time distribution of events.

In this article we will deal only with the time variation of the exposure as this 
is usually the most delicate issue. Our method can be 
applied to any kind of detector with nearly time independent aperture, a typical 
property of ground arrays where the detector has, in principle, 100\% duty cycle. 
Through the text, just for illustration purposes, we are going to work with a cosmic ray 
detector located at the same position of the Auger Observatory \cite{Abraham:2004dt}, but 
the results should be valid for a large variety of detectors.

The paper is organised as follows: we present the basic underlying assumption of 
Julian day $\times$ solar factorization of the detector rate in section \ref{assumption}, 
and in section \ref{getRfromdata} we show how to recover the time detector distribution from 
the dataset itself if the possible sky anisotropies, through their sidereal amplitudes, 
are no larger than just a few percent, therefore avoiding the use of Monte Carlo simulation. 
The residual systematic effects in the coverage due to possible non-factorizable 
components of the detector rate is estimated in section \ref{residual} by using 
a model for the detector rate based on weather monitoring data taken at the Auger 
site. The robustness of coverage maps built under the factorization hypothesis is 
treated in section \ref{perform} by reconstructing some anisotropic patterns 
on the sky. We finally conclude in section \ref{conclu}.

\section{The Factorizable Acceptance Model (FAM)}
\label{assumption}

Let ${\cal A}$ be the detection efficiency of a surface array. This function depends 
on the time 
$t$, on the shower horizontal coordinates $\theta$ and $\phi$ and on a set 
of parameters, like the energy, which we will refer to as {\bf X}. By writing the 
time as a function of two distinct variables: $t=t(j,s)$, where $j$ is the Julian day 
and $s$ is the solar time, we can therefore represent the array efficiency as 
${\cal A}(j,s,\theta,\phi,\mathbf{X})$.

In the most general case, the efficiency will have an intricate form with non-trivial 
correlations among all the parameters. For example, at a fixed energy, the weather 
induced time modulations might have a relative amplitude which depends on the zenith 
angle. In the following, one assumes that the domain spanned by the variables 
$\{\theta,\phi,\mathbf{X}\}$ can be broken up into complementary regions over which 
${\cal A}$ can be factorized as 
\begin{equation}
{\cal A}(j,s,\theta,\phi,\mathbf{X}) = {\cal A}_{t}(j,s) \times 
{\cal A}_{\Omega}(\theta,\phi,\mathbf{X}),
\label{fact1}
\end{equation}
where ${\cal A}_{t}(j,s)$ and ${\cal A}_{\Omega}(\theta,\phi,\mathbf{X})$ are both 
dimensionless, and the results are given for only one of such 
regions. In practice, to get the full picture, one does the same analysis over all 
regions required to be defined.

In the factorizable acceptance model (FAM), we assume that the time variation of the 
efficiency is the product of a daily variation in solar time $\eta_{D}(s)$ and a 
seasonal variation in Julian days $\eta_{Y}(j)$, so that one can write
\begin{equation}
{\cal A}_{t}(j,s) = \eta_{Y}(j) \times \eta_{D}(s).
\label{fact2}
\end{equation}

The hypothesis behind this factorization is that what happens during a day is not 
exclusive to that particular day, but mainly correlated to the sun altitude in the sky 
(given by the solar time) in that day. The validity of such a hypothesis can be 
checked directly from the data set as will be shown in section \ref{residual}.

\section{Extracting the acceptance under the FAM hypothesis from the 
data set itself}
\label{getRfromdata}

The goal of this section is to show how one can retrieve the time modulations of 
the detector acceptance from the data set itself, if one uses an integer number 
of data taking years and if a possible anisotropic pattern 
on the sky has a small amplitude, that is, a sidereal time modulation no larger than 
a few percent. For ultra high energy cosmic rays, this is in fact a pretty good 
assumption\footnote{\label{argument}We are assuming here as ultra high energy 
cosmic rays those with energies above $\sim 10^{15}$ eV. Therefore, the statement 
is based on the isotropy at PeV energies observed 
by KASCADE \cite{Antoni:2003jm}, the Auger results \cite{ICRCEric}, and it remains 
valid even if we take the AGASA dipole \cite{Agasadipole} as real.}.

Let ${\cal F}(\alpha,\delta,\mathbf{X})$ be the cosmic ray flux from the direction 
$(\alpha,\delta)$ in equatorial coordinates, where $\alpha$ is the right ascension and 
$\delta$ is the declination, reaching the Earth per unit time, area, solid 
angle and per units of $\mathbf{X}$. The variables $\alpha$ and $\delta$ can be 
written in terms of the local horizontal coordinates $\theta$, $\phi$ and the time 
of detection $t(j,s)$. Therefore, the following integral over one of the regions where 
the factorization (\ref{fact1}) is valid
\begin{equation}
\Phi(t) = \int\limits_{\mathbf{X}}\int\limits_{\Omega}
{\cal A}_{\Omega}(\theta,\phi,\mathbf{X})
{\cal F}(\alpha(t,\theta,\phi),\delta(t,\theta,\phi),\mathbf{X})
d\mathbf{X}d\Omega,
\end{equation}
has dimension of number of particles per unit time and area. The rate of cosmic rays 
detected (over the region where $\Phi(t)$ is defined) as a function of time is then given by
\begin{equation}
{\cal R}(t) = {\cal R}(j,s) = {\cal A}_{t}(j,s) \times \Phi(t) \times 
S_{eff}(t), 
\end{equation}
where $S_{eff}(t)$ represents here the instantaneous effective area of the detector, since 
this might not be constant in time, as is the case of detectors which take data as their 
sizes are still growing by the installation of new detection units, or experimental 
problems which might black part of the detector, reducing its effective area. These 
effects can usually be corrected for and we will not be concerned with them in this paper.

The function $\Phi(t)$ gives us the time dependence of the sky anisotropy and can be written 
in terms of the sidereal time $t_{sid}$ as
\begin{equation}
\Phi(t_{sid}) = \Phi_{0}\left[1+\epsilon(t_{sid})\right],
\end{equation}
where $\epsilon(t_{sid})$ is a function of period $D_{sid}$ (a sidereal day) with null 
integral over this time interval and $|\epsilon| << 1$ and
\begin{equation}
\Phi_{0} = \frac{1}{D_{sid}}\int\limits_{D_{sid}}\Phi(t_{sid})dt_{sid},
\end{equation}
Since a Julian 
day is about only 4 minutes longer than a sidereal day (a correction factor of about 
1 + 1/365.25) 
and a calendar year being a little longer than the time it takes for $t_{sid}$ to 
span the whole sky in right ascension (same correction), we have two useful 
approximations
\begin{equation}
\int\limits_{Y_{sol}}\epsilon(t_{sid}) dt_{sid} \simeq 0, \quad \textrm{and} \quad 
\int\limits_{D_{sol}}\epsilon(t_{sid}) dt_{sid} \simeq 0,
\label{symmetry}
\end{equation}
where $Y_{sol}$ and $D_{sol}$ are the durations of a solar year and a solar day, 
respectively.

By explicitly writing the average values of $\eta_{Y}(j)$ and $\eta_{D}(s)$, we have
\begin{equation}
\eta_{Y}(j) = \overline{\eta}_{Y}\left(1+f_{Y}(j)\right) \quad \textrm{and} \quad
\eta_{D}(s) = \overline{\eta}_{D}\left(1+f_{D}(s)\right),
\end{equation}
and by definition the fractional amplitudes satisfy
\begin{equation}
\int\limits_{Y_{sol}}f_{Y}(j) dj = 0, \quad \textrm{and} \quad 
\int\limits_{D_{sol}}f_{D}(s) ds = 0,
\label{symm_exact}
\end{equation}

With all that in mind, we can write the rate of cosmic rays, corrected by the 
effective array surface as
\begin{eqnarray}
{\cal R}^{'}(j,s) &=& \frac{{\cal R}(j,s)}{S_{eff}(j,s)} \nonumber\\
&=&\overline{\eta}_{Y}
\overline{\eta}_{D}
\Phi_{0}\left[1+\epsilon(t_{sid})\right]\left[1+f_{Y}+f_{D}+f_{Y}f_{D}\right],
\end{eqnarray}
where we have omitted the dependence of $f_{Y}$ and $f_{D}$ on $j$ and $s$, respectively. 
Integrating over a solar day or a solar year and using eqs. (\ref{symmetry}) and 
(\ref{symm_exact}), we obtain the functions 
${\cal R}_{Y}^{'}(j)=\int\limits_{D_{sol}}{\cal R'}(j,s)ds$ and 
${\cal R}_{D}^{'}(s)=\int\limits_{Y_{sol}}{\cal R'}(j,s)dj$
\begin{equation}
{\cal R}_{Y}^{'}(j)=
\Phi_{0}\overline{\eta}_{D}D_{sol}\left[1+\frac{1}{D_{sol}}
\int\limits_{D_{sol}}\epsilon(t_{sid})f_{D}(s)ds\right]\eta_{Y}(j)
\end{equation}
\begin{equation}
{\cal R}_{D}^{'}(s)=
\Phi_{0}\overline{\eta}_{Y}Y_{sol}\left[1+\frac{1}{Y_{sol}}
\int\limits_{Y_{sol}}\epsilon(t_{sid})f_{Y}(j)dj\right]\eta_{D}(s).
\end{equation}

We are specifically interested in the estimation of the coverage map for cosmic rays at 
energies above around $10^{17}$~eV, detected by cosmic rays detectors such as large surface 
arrays. As already mentioned, such arrays are quite 
stable in time, with typical fractional amplitudes for the time variations in the detection 
efficiency which are no larger than 10\% (a total 20\% variation) along a day or a year (see 
section \ref{residual}). 
Moreover, the sidereal time modulation at these energies will not exceed the 5\% level, a 
somewhat 
exaggerated scenario (see footnote on page \pageref{argument}). Under these hypotheses the 
integral inside square brackets can be neglected, since it will be certainly 
below the 0.5\% level.

Therefore, one can write 
\begin{equation}
{\cal R}_{Y}^{'}(j) \simeq \Phi_{0}\overline{\eta}_{D}D_{sol}\eta_{Y}(j), 
\quad
{\cal R}_{D}^{'}(s) \simeq 
\Phi_{0}\overline{\eta}_{Y}Y_{sol}\eta_{D}(s),
\label{aprox_final}
\end{equation}
which contain only the detector efficiency functions, regardless of any anisotropy 
on the sky. The left hand side of these two equations are distributions which can be easily 
obtained from the data set itself as the histograms of physical events as a function 
of the Julian day or the solar time. It is therefore possible to extract the detector 
information from the data, even in the presence of anisotropy. The approximations in 
eq. (\ref{aprox_final}) are good up to order $|\epsilon||f_{Y(D)}|$ (the product of the absolute 
values of $\epsilon$ and $f_{Y(D)}$). For all those purposes 
where the required precision over the coverage estimation lies below this level, the 
procedure outlined here does not apply.

A coverage map can then be built from the data by using the local reconstructed angles 
$\theta$ and $\phi$ and sorting the arrival times according to the quantity 
\begin{equation}
{\cal R}(t)/\Phi(t) \propto {\cal R}_{Y}^{'}(j) \times {\cal R}_{D}^{'}(s) \times S_{eff}(j,s),
\label{fact3}
\end{equation}
where the proportionality factor depends on $\Phi_{0}$, $\overline{\eta}_{Y}$ and 
$\overline{\eta}_{D}$, being therefore unknown, but completely unimportant, since 
it is a global constant factor not depending on time.

\section{The residual systematic due to non-factorizable components in the 
acceptance}
\label{residual}

The trigger rate of a surface array depends on the weather conditions, since the 
changes in the pressure $P$ and air density $\rho$ affect the shower 
development through the atmosphere. An increase in air density will reduce the Moli\`ere 
radius, whereas the average shower age at ground increases with the atmospheric pressure. 
As shown in \cite{Bleve:2007mb}, 
the Auger surface detector trigger rate can be fairly well modelled through a linear 
dependence with $P$ and $\rho$. A fit to the data collected by the observatory spanning a 
2-year period have produced:
\begin{equation}
{\cal R}(t) = R_{0}\left[1 + a_{P}(P-P_{0}) + a_{\rho}(\rho_{d}-\rho_{0})+b_{\rho}(\rho-\rho_{d})
\right],
\label{fitrate}
\end{equation} 
where $\rho_{0}=1.055$ kg m$^{-3}$ is the average reference density, $\rho_{d}$ is the 
average daily density, $a_{P}=(9\pm5)\times 10^{-4}$ hPa$^{-1}$, 
$a_{\rho}=(-2.68\pm 0.07)$ kg$^{-1}$m$^{3}$ and $b_{\rho}=(-0.85\pm 0.07)$ kg$^{-1}$m$^{3}$.
\begin{figure}
\vglue -0.6cm
\center{
\includegraphics[width=3.8in, angle=0]{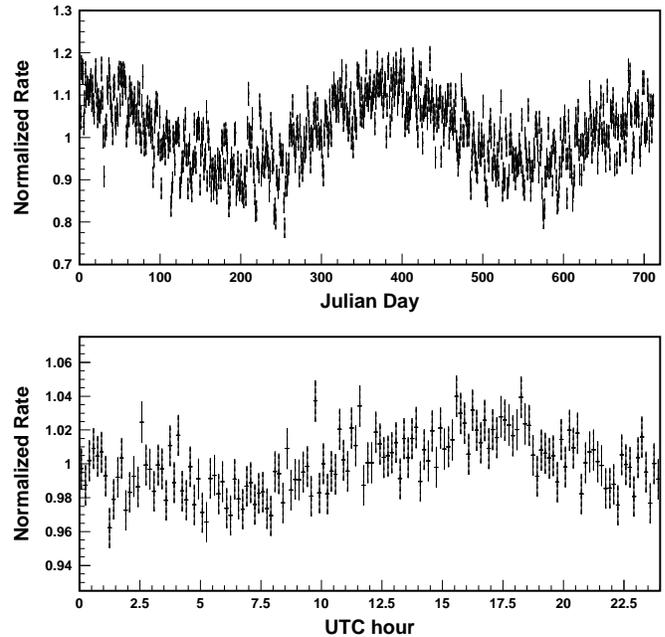}
}
\vglue -0.6cm
\caption{Seasonal (top) and daily (bottom) rate variations according to the model represented 
by equation (\ref{fitrate}) shown in ref. \cite{Bleve:2007mb} to describe fairly well the rates 
observed in the Auger data set. The rate is assumed to depend linearly on the ground pressure 
and air density (see coefficients in the text and \cite{Bleve:2007mb}). The latter is obtained 
from the temperature. Pressure and temperature measurements were taken from a weather station 
on site and two years of data are shown: 2005-2006.}
\label{rate_jdst}
\end{figure}

From now on, we will assume that the trigger rate of a surface array is well modelled by 
eq. (\ref{fitrate}), representing appropriately both its long term seasonal and daily 
modulations, as well as its short time fluctuations due to the underlying variations in 
pressure and density\footnote{Here, we assume the air density is completely determined by 
$P$ and $T$ according to $\rho=p/R_{dry}T$, where $R_{dry}=287.05$ J/kg/K is the specific 
gas constant for dry air, that is, we neglect the effect of humidity.}. Everywhere in the 
paper, we assume that the array is located at the same geographical position as the Auger 
detector (35.25$^\circ$ S, 69.25$^\circ$ W). Figure \ref{rate_jdst} 
shows the rate predicted by eq. (\ref{fitrate}), using the same weather monitoring data of 
reference \cite{Bleve:2007mb}, as a function of the Julian day and the solar time. Two years 
are shown (2005-2006) and one can see the $\sim$6\% seasonal modulation (top) and daily 
(bottom) variation of $\sim$2\%.

In order to estimate the possible residual systematic due to the presence of components in 
the trigger efficiency whose time behaviour cannot be factorized into a seasonal and a diurnal 
effect, we simulated an isotropic cosmic ray distribution, convolving it with a detector 
time efficiency given by eq. (\ref{fitrate}) and a simple geometrical efficiency given by 
$\sin\theta\cos\theta$, which takes into account both the solid angle and the effective 
detector transversal area seen by a shower at zenith\footnote{For a real 
surface array, there are several additional effects which contribute to the coverage 
estimation. However, as already stated before, we are only interested here in the role of 
the time behaviour of the trigger efficiency.} angle $\theta$. All the simulations in this 
paper include only 
showers with $\theta<60$ degrees, consistent with the data sample used in \cite{Bleve:2007mb}. 
In the absence of 
anisotropic patterns from the sky, we know that all distortions observed in the arrival 
time distribution of the showers are due only to local effects. Therefore, the true 
coverage map can be built directly from the data set by performing a scrambling of the 
arrival times. The exposure sky map built in this way will be denominated here UTC, 
since the arrival times are usually given as UTC time.
\begin{figure}
\vglue -0.6cm
\center{
\includegraphics[width=3.5in, angle=0]{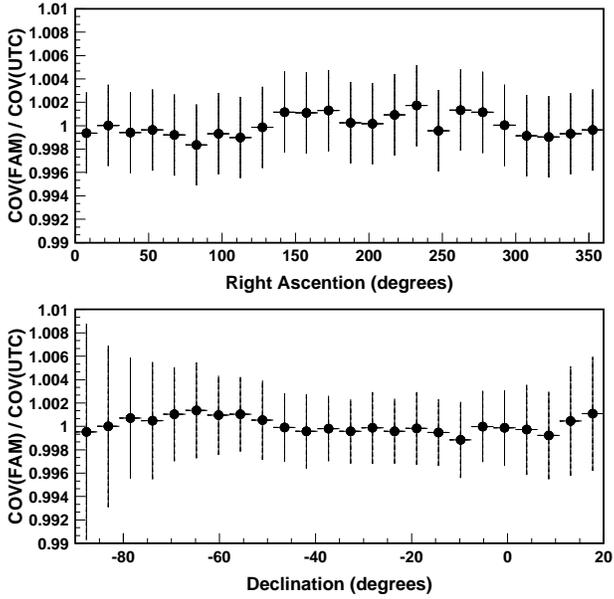}
}
\vglue -0.6cm
\caption{Ratio between the FAM and the UTC sky coverage map as a function of right 
ascension (top) and declination (bottom). The exposure maps were estimated by using 
simulated data with events following the pressure and temperature modulated rate modelled 
by eq. (\ref{fitrate}). The error bars represent the statistical error only. Here, since 
there is no anisotropic pattern from the sky, the true coverage is represented by the UTC 
sky map.}
\label{cov_ratio}
\end{figure}

The validity of the factorization hypothesis can then be tested by comparing the UTC 
coverage with the one obtained by replacing the original arrival times by a value drawn 
according to eq. (\ref{fact3}). The latter map will be called the FAM coverage. The 
coverage ratio FAM/UTC is shown in figure \ref{cov_ratio} binned in right ascension 
and declination. To beat the statistical fluctuations down and have access to small 
systematic effects, these plots are based on large Monte Carlo samples ($8\times 10^{6}$ 
events). One can see that the residual modulations are below the 0.2\% level and 
are still consistent with the statistical fluctuations. Therefore, the coverage map of 
a surface array whose time trigger efficiency can be fairly well described by a linear 
dependence on air pressure and density as given by eq. (\ref{fitrate}), can 
be accurately (that is, up to 0.2\% level) built under the hypothesis of seasonal $\times$ 
diurnal effects factorization.

\section{The method performance under large scale anisotropy patterns}
\label{perform}

Having validated the hypothesis of factorization in the last section, one can study 
the performance of the FAM and UTC coverages when in addition to local effects, 
the arrival time distribution is further distorted by large scale anisotropy 
patterns from the sky.

 A coverage map constructed from the data set itself 
which keeps the original arrival time distribution, like the UTC coverage, which essentially 
makes a scrambling of the arrival times, will always underestimate the amplitude of a right 
ascension modulation, since all of the variation induced into the time distribution is 
absorbed into the coverage. The usual excess maps made by taking the signal/coverage ratio 
will therefore underestimate the modulation in right ascension.
\begin{figure}
\vglue -0.6cm
\center{
\includegraphics[width=3.5in, angle=0]{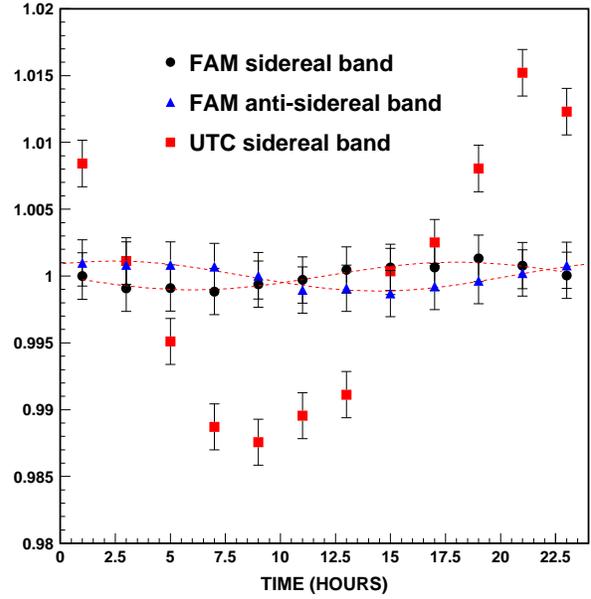}
}
\vglue -0.6cm
\caption{Sidereal modulation for the UTC coverage (red squares) and the sidereal (black circles) 
and anti-sidereal (blue triangles) sidebands for the FAM coverage. The equal amplitudes sidereal 
and anti-sidereal sidebands for the FAM coverage are the results of the seasonal modulated solar 
variation. The dotted lines are fits to the respective sidebands with amplitudes: 
$(0.10 \pm 0.07)$\% (sidereal) and $(0.11 \pm 0.07)$\% (anti-sidereal). We have artificially 
introduced a shift in the phase of the anti-sidereal band for better visualisation, since this 
one is practically in phase with the sidereal band.}
\label{sid_asid_rec}
\end{figure}

Figure \ref{sid_asid_rec} shows the time distribution used to build the coverage under the FAM 
hypothesis and the usual scrambling of UTC shower arrival time, when an equatorial dipole of 
amplitude 2\% is present. The dipolar pattern induces a large sidereal contamination into 
the UTC coverage as can be seen in figure \ref{sid_asid_rec}. In fact, this is one of the most 
unfavorable cases for a UTC like coverage map, since it is a pure right ascension distortion, 
creating a large modulation in the events arrival time distribution. The concept of anti-sidereal 
time was introduced some time 
ago by Farley and Storey \cite{FS} in order to study long term behaviour of cosmic ray detectors. 
The authors made use of the fact that a seasonal modulated diurnal variation (both assumed to 
be harmonic) could be obtained 
by the interference of sidebands of equal amplitudes and slightly different periods, in which 
the slightly higher frequency (with respect to a solar frequency) sideband is the well known 
sidereal band, whereas the slightly lower frequency sideband was called the anti-sidereal 
component\footnote{An anti-sidereal 
year has approximately one day more than a solar year and a sidereal year has one day less.}. 
Therefore, even pure local seasonal and diurnal modulations will give rise to an apparent 
sidereal anisotropy, but these can then be identified by looking for the associated 
equal amplitude anti-sidereal sideband. The time distribution of events used in the FAM 
coverage shows in fact such twin sidebands with fitted amplitudes of about 0.1\%. The amplitude 
of the sidebands is given approximately by the product of the amplitudes of the seasonal and 
the diurnal components. The fitted amplitudes to the rates of figure~\ref{rate_jdst} imply an 
expected 6\% $\times$ 2\% = 0.12\% sidereal (anti-sidereal) modulation, which is in pretty good 
agreement with figure \ref{sid_asid_rec}. Therefore, by assuming a Julian day $\times$ solar time 
factorization of the rate, essentially none of the 
right ascension modulation leaks into the final exposure sky map. It is worth stressing, though, 
that such a cancellation of sidereal contribution from sky anisotropy can only be achieved by 
using an integer number of years as already discussed in section \ref{getRfromdata}. We stress as 
well that unlike Farley and Storey which assume a harmonic form for the seasonal and 
diurnal modulations, in the FAM method, their analytical forms are extracted directly from the 
data.

\begin{figure}
\center{
\includegraphics[width=3.5in, angle=0]{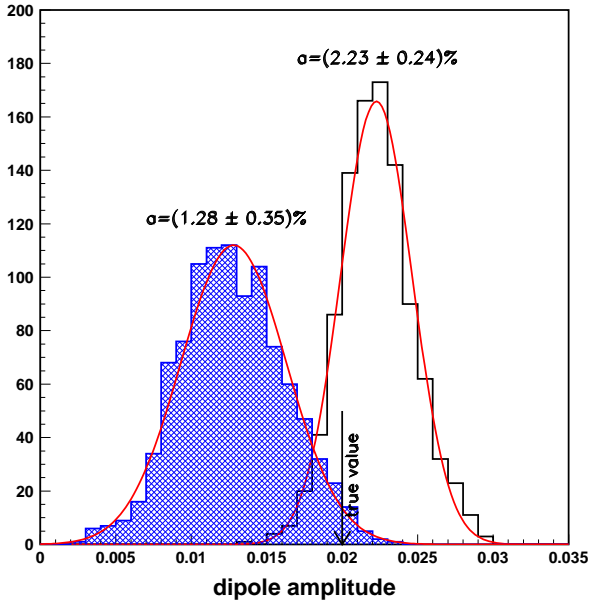}
}
\vglue -0.6cm
\caption{Distribution of reconstructed amplitudes of an equatorial 2\% amplitude dipole via 
the method of reconstruction of pseudo-multipolar coefficients for non-uniform and incomplete 
sky coverage \cite{Billoir:2007kb}. Two different coverages: 
UTC (shaded blue) and FAM, and used as input to the reconstruction are compared. The central 
values and the standard 
deviations for the fitted Gaussians are shown on the top of each histogram. Each one of the 
1000 samples shown have $5\times 10^5$ events.}
\label{dip_equ_rec}
\end{figure}

A more quantitative idea of the bias introduced into the coverage by a simple scrambling of the 
original arrival times is given in figure \ref{dip_equ_rec}. The plot shows the distribution 
of reconstructed amplitudes for 1000 realisations of the 2\% equatorial dipole. The reconstruction 
of dipolar patterns with a full sky observatory was first analysed in \cite{Sommers:2000us} with 
later generalisations to the case of partial sky coverage presented in \cite{Aublin:2005nv} and 
\cite{Mollerach:2005dv}. In figure \ref{dip_equ_rec}, the reconstruction 
is performed by retrieving the set of multipolar coefficients of the events directions map for a 
non-uniform and, in this case, even incomplete sky exposure. The central idea of the 
method is that the {\it pseudo} multipolar coefficients obtained with non-uniform partial sky 
coverage are related to the true ones by a convolution operation whose kernel is a function of the 
detector window function (the coverage map) \cite{Billoir:2007kb}.  

From figure \ref{dip_equ_rec} we see a clear underestimation of the dipole amplitude when the 
reconstruction is done by 
using the UTC sky map, with the average value ($1.28\pm 0.35$)\% being 2$\sigma$ 
away from the true value. With the FAM coverage, in turn, one is able to retrieve the 
amplitude within 1$\sigma$. The relative statistical uncertainty of 11\% in the amplitude for 
the FAM case is also smaller than the 27\% corresponding to the UTC coverage.

It is interesting to compare these coverage based reconstructions with a coverage independent 
one. For a pure right ascension (that is sidereal) modulation, it is straightforward to estimate 
the sidereal amplitude through the power distribution at the sidereal frequency of the arrival 
time Fourier transform as described in \cite{Billoir:2007nu}. The reconstruction is done in a two step 
way. Firstly, the normalised Fourier coefficients of a set of $N$ modified times 
$\{t_{i}\}$\footnote{It was shown in \cite{Billoir:2007nu} that by correcting the arrival times by the 
right ascension phase of the event with respect to the local sidereal time, one is able to 
observe modulations on smaller scales and resolve, for example, the sidereal and diurnal 
frequencies as long as data is gathered for a sufficiently long period of time. All the 
simulations in this section are for a 5-year period.} are calculated at frequency $f$
($\omega=2\pi f$) as $a(f)=c(f)+is(f)$
\begin{equation}
c(f) = \frac{2}{W}\sum_{i}w_{i}\cos(\omega t_{i}), \quad 
s(f) = \frac{2}{W}\sum_{i}w_{i}\sin(\omega t_{i})
\label{fourier}
\end{equation}
with $W=\sum_{i}w_{i}$ ($w_{i}=1$), where the sums run all over the $N$ events. An estimation for the 
annual (diurnal) amplitude 
$p_{y(d)}=\sqrt{s^{2}(f_{y(d)})+c^{2}(f_{y(d)})}$ and phase 
$\phi_{y(d)}=\textrm{atan}(s(f_{y(d)}),c(f_{y(d)}))$ is then obtained. Finally, the Fourier amplitudes 
in (\ref{fourier}) are recalculated, now with weights 
\begin{equation}
w_{i} = \left[(1+p_{y}\cos(\omega t_{i} - \phi_{y}))\times(1+p_{d}\cos(\omega t_{i} - \phi_{d}))\right]^{-1}
\label{weights}
\end{equation}
to get the final sidereal amplitude and phase. Such a Fourier analysis, totally independent of 
the coverage map, gives an amplitude of $(1.6 \pm 0.2)$\% for the same 2\% equatorial dipole, 
a result somewhere in between the reconstructions through the multipolar coefficients. 
\begin{figure}
\vglue -1.5cm
\center{
\includegraphics[width=3.5in, angle=0]{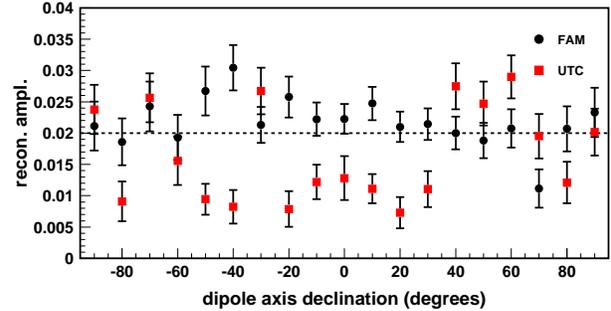}
}
\vglue -3.6cm
\caption{Dipole amplitudes reconstructed through the deconvolution of the multipolar coefficients 
for non-uniform and incomplete sky coverage \cite{Billoir:2007kb} as a function 
of the declination of the dipole axis. FAM (black circles) and UTC (red squares) based 
coverage maps are compared. The true 2\% amplitude is represented by the horizontal dashed 
line.}
\label{dip_dec_rec}
\end{figure}

A more systematic test of the performance of the FAM and UTC based coverage maps is shown 
in figure \ref{dip_dec_rec}, where the reconstructed amplitudes and their associated errors 
are shown for dipoles with axis at different declinations (and fixed amplitude of 2\%). By 
varying the angle between its axis and the equatorial plane, one changes the amplitude of 
the induced sidereal variations. One can see that, in average, the amplitude is more accurately 
retrieved when the FAM coverage map is used as input to the multipolar reconstruction. In the 
whole range between -90$^{\circ}$ and 90$^\circ$, the UTC coverage leads to a mean amplitude 
2.9$\sigma$ away from the true value, whereas a 1.3$\sigma$ estimation is achieved with the 
FAM sky map.

\section{Conclusions}
\label{conclu}

We have introduced a new method for estimating the exposure sky map of a cosmic ray 
detector, based on the assumption of factorization of the time behaviour of such a detector 
into seasonal (represented by the Julian day) and diurnal (represented by the solar 
time) variations. Such a hypothesis is motivated by the current knowledge on the modulations 
of the detection rate of large surface arrays caused by weather effects, such as pressure and 
air temperature, which affects both the longitudinal and lateral developments of particle 
cascades through the atmosphere. By using an 
integer number of data taking years, one is able to cancel the sidereal contribution of sky 
anisotropy to the Julian day and solar time histrograms, allowing to extract the detector 
time behaviour from the data set itself.

In \cite{Bleve:2007mb}, a model was shown to describe fairly well the yearly and diurnal 
rate modulations observed in the Auger data set, where the rate is assumed to vary 
linearly with pressure and air density, with the proportionality coefficients being 
fitted directly to the number of showers detected by the surface array and to the 
atmospheric monitoring data taken regularly on site. Using this model, we have shown 
that possible residual systematic effects caused by the presence of non-factorizable 
terms into the long term detector behaviour will induce fake large scale anisotropy 
modulations on the sky no larger than 0.2\%, being therefore negligible for a typical 
1\% desired accuracy into the coverage estimation.

Coverage maps built under the factorization hypothesis were shown to be more suitable 
when reconstructing dipolar anisotropic patterns through the recovery of its multipolar 
coefficients, providing amplitudes with a mean systematic shift of 1.3$\sigma$ with 
respect to the true value as one changes the declination of the dipole axis. The amplitudes 
reconstructed from a coverage where one simply scrambles the showers arrival times are 
strongly biased, with values tipically around 3$\sigma$ away from the true values. A 
coverage independent 
estimation of the equatorial dipole, such as the one provided by the Fourier analysis 
of the arrival times and detector running for a 5-year period gives an amplitude 
somewhere in between the multipolar method with UTC and FAM input coverages, with a 
systematic underestimation of the amplitude at the $2\sigma$ level.

There are some similarities behind the factorization hypothesis with the analysis performed 
in \cite{FS}, but whereas Farley and Storey assume a particular harmonic dependence for 
both the seasonal and diurnal modulations in order to fit their respective amplitudes, the 
time distribution used to build the FAM coverage is taken from the data set itself.

\section*{Acknowledgements}

This work was supported by the Conselho Nacional de Desenvolvimento Cient\'ifico e 
Tecnol\'ogico (CNPq), Brazil, and the Centre National de la Recherche Scientifique, 
Institute National de Physique Nucl\'eaire et Physique des Particules (IN2P3/CNRS), 
France. LEO - International Research Group. The authors are grateful to the Pierre 
Auger Collaboration for kindly supplying the monitoring data used in this work and 
for fruitful discussions with some of its members. The authors thank particularly 
O. Deligny for the help with the multipolar reconstruction.

\end{document}